# Model for heterogeneous reaction-diffusion systems with application to one epidemic

Orlando Silva

Departamento de Física, Universidade de Évora, Portugal

orlando@uevora.pt

**Abstract**

The dynamics of ecological as well as chemical systems may depend on heterogeneous configurations. Heterogeneity in reaction-diffusion systems often increase modelling and simulating difficulties when non-linear effects are present. One synthetic epidemic system with short range heterogeneous composition is modelled and its space-time evolution studied using maximum heterogeneity details. Two other modelling alternatives are applied, one of them using elementary mean-field variables, one other using non-localized geometrical parameters, so avoiding the limitations of the used mean-field model, while keeping significant features of more detailed models. Both the detailed and the mean-field models are solved by means of the standard finite volume method. The model with less defined geometry is solved by means of one modified version of the finite volume method. Simulation results of the three models are compared. At the high diffusion range all models behave similarly. At moderate diffusion fluxes, the numerical results of the model with reduced geometric details are in excellent agreement with the results of the detailed model. The simple mean-field model presents limited accuracy at low and moderate values of the diffusion coefficient.

## 1. Introduction

Transport and reaction phenomena occur in numerous systems of various kinds, categorized as continuous or discrete, solid or fluid, composed by chemical, biologic, or other elements, with varied relations between the elements, various types and levels of heterogeneity, in an unrestrained number of classifications of the types of systems where transport and reaction can happen. Solid and fluid mixtures may have reaction and diffusion of heat and chemical species, like in forest fires. Similar phenomena occur in ecological systems, in population dynamics and epidemics. These are some examples of heterogeneous systems in which reaction and diffusive transport may be recognized. Other forms of transport besides diffusion are not taken in account here.

The heterogeneities may be of compositional nature, different locations of the system having different composition. Other kinds of heterogeneities may occur affecting the behaviour of the elements, with non-uniform rules affecting parameters of the system, originated by internal or external actions, either perfectly known or not completely defined.

A system with heterogenic composition is defined and studied in which reaction and diffusion occur. An epidemic system involving only two epidemic species, susceptible and infected, will be proposed in which the susceptible density S presents initial short range heterogeneity. Non-linear reaction rates between species are present. The infected I density is initially zero everywhere except in one small region. The system is described as one simplified form of the systems in Noble 1974 and Silva 2016, to which heterogeneous composition is added.

One system is usually defined as one part of the universe whose elements have mutual interactions, external interactions also possibly being important. One system to be studied must in principle be defined with so many specifications as possible, the main elements by which is it composed, the domains where they are defined, the relations among the elements and with as many interactive objects as possible, internal or external. Often the system to be studied is complex and the definition of some of the main elements and important relations are not available.

The cases are not frequent in which one real system is thoroughly specified in its elements, with completely defined rules of behaviour and relations with all other internal or external elements. One perfect mathematical model may then be built as to describe the structure and evolution of the system.

However, even when all the elements and relations are manageable, the modeller and the user may prefer to use simpler models that do not include all the known details of the system. This may happen in cases of difficult access to data, by limitations of the available mathematical simulation methods, or by any other reason. The model definition then ceases to precisely depict the complete system.

Heterogeneous distributions of reacting elements may originate large spatial variations of reaction rates and consequently of diffusion fluxes. When the reactive or contagion source term is non-linear, the infection rates normally alter if the original heterogeneous S field is replaced by one less detailed field. Detailed description of heterogeneous fields is however not always available or easy to work with. So, the need exists of making simple and workable models of the heterogeneous characteristics of the systems.

The objective of the study consists in describing and simulating the evolution of the reaction-diffusion system, one SI epidemic system. Three different models are defined, numbered in growing order of detail, Model 3 presenting the maximum detail of the heterogeneity of the chosen system. The simplest Model 1 replaces short range heterogeneity by short range averages, some non-linear effects being lost. Model 2 incorporates short range heterogeneity with less detail than Model 3, responding to the possible unavailability or excessively demanding description of heterogeneities.

Alternatives to the use of detailed heterogeneity descriptions and to the use of mean field variables exist. The use of moment closure models to include the effects of short range heterogeneity is addressed in Christakos et al. 2005, Parham et al. 2006, and references there.

Heterogeneity may be present in initial values of dependent variables, in space or time variations of the values of properties (see for example Kinezaki et al. 2003). Heterogeneity may also be associated with changes of external conditions and with heterogeneous variation of some of the system general rules.

By means of continuous analytical reasoning, Alonso et al. 2009 derive effective parameters for reaction-diffusion equations, then applying to reaction-diffusion systems with several types of heterogeneous parameters. Kinezaki et al. 2003, studied the invasion dynamics of a single species in a regularly striped environment, a two-dimensional extended Fisher equation was numerically solved, obtaining the spatial-temporal population spread applied to the chosen spatial heterogeneity configuration. This work followed a previous work of Shigesada et al. 1986 that studied an analogous one-dimensional system. The spontaneous appearance of spatial heterogeneities is reported in the work of Petrovskii et al. 2005. They studied the dynamics of a predator-prey system, concluding for the generality of the patch phenomenon and conditions of occurrence. From extensive data analysis of measles epidemics, Grenfell et al. 2001 reassured the existence and regularity of travelling waves. Various mathematical methods and models exist to deal with systems of biologic species, Johnson et al. 2006, Parham and Ferguson 2006, Hastings et al. 2005, referring several other studies on models and the effects of spatial heterogeneity. Lund et al. 2013, studied the effects of density heterogeneity on one epidemic spreading by means of reaction-diffusion in networks.

The dynamics of the epidemic system is here described by differential balance equations of susceptible and infected populations. To solve these equations, the finite volume method will be used. Three different heterogeneity models will be inserted in the algebraic equations of finite volume method used to simulate the epidemic system evolution.

The simulations of the evolution of the system according to Models 1 and 3 will be made by means of the classic finite volume method. One modified version of FVM will be used to simulate the space and time evolution by Model 2, the detailed heterogeneous S field being replaced by one less detailed field, yet keeping important geometric features.

Model 1 uses mean-field variables replacing short-range heterogeneities by spatial averages of the heterogeneous quantities. This form of the mean-field equations thus erases the effects of short range heterogeneities.

Model 2 replaces the detailed description of heterogeneous fields by a description in which some geometrical details are replaced by artificial quantifications of the sizes, mutual contact areas and distances. These specifications may be guessed or determined from the existing knowledge about the system heterogeneities.

In Model 3 the system is described by the most detailed description of the heterogeneous fields. The simulation results of this Model, used as reference, will be compared with the two other Models' results.

Simulations of Models 1 and 3 use the classical finite volume method. In the case of Model 2 a modified version of the finite volume method is used. The use of numerical methods instead of searching analytical solutions provides the possibility of getting results for a vast number of regular or irregular conditions. Moreover it is possible to insert convection and long range transport, attractive potentials, Murray 1989, as well as stochastic calculations in the originally deterministic balance equations.

From the simulations using the three Models, a number of results will be obtained that display the behaviour of the system as a function of the Models used in calculations. Although the methods are applied to one chosen artificial heterogeneous epidemics, they may be applied to real epidemics.

Section 2 defines the system and three Models to describe it. The system balance differential equations are discretized to obtain the balance equations of the three Models. The methods used in the simulations of the evolution of the system according to the Models are described. Section 3 displays the results of the space-time evolution of density I; the evolution in time of the total numbers of infected and susceptible at different diffusion coefficients D; the final numbers of susceptible at various D calculated by the three Models; the final S fields at different D calculated by Model 3. Some interpretations of the results and conclusions are presented in sections 3 and 4.

## 2. System, models and methods

### 2.1. The system

The arbitrated system to be studied in the present work is one epidemic in which the susceptible and infected species have non-uniform initial distribution. The system is supposed to exist in one rectangular domain with sides of 300 and 150 arbitrary length units divided in 237x117 rectangles. In both directions, one in 3 rectangles has the initial value of the density of susceptible species S=100. The remaining rectangles have initial S=20. The initial density of infected is I=0 everywhere excepting in the rectangle approximately centred at (150,1.923), where I=1 and S=100.

In this SI epidemic system the contagion rate only depends on the instantaneous densities of infected and susceptible. Only the infected are supposed to move, and the mortality is supposed to only depend on the instantaneous local number of infected. The system is considered isolated, no interactions exist with the outside of the domain.

This system is modelled in a similar way to the one in Silva 2016, with a few simplifications. The variation rate of the local density of susceptible S is supposed to depend only on the contagion rate assumed proportional to the densities of susceptible and infected. The variation rate of the local density of infected I is calculated by adding the rate of formation of infected density by contagion, the mortality rate density and the infected net flux density. The flux density of infected between two locations is supposed to be proportional to the gradient of I.

The basic rules of the epidemic dynamics can then be expressed by means of Eq.s 1-2, expressing the balance of the epidemic species in terms of the source rates and fluxes of the densities S and I.

$$\frac{\partial I}{\partial t} = k_c SI - \mu I + \nabla . D \nabla I \tag{1}$$

$$\frac{\partial S}{\partial t} = -k_c SI \tag{2}$$

The system is supposed to have no interactions with the external world, and so S and I have zero flux at the outer boundaries of the rectangular domain in the directions normal to these boundaries.

The coefficients of contagion and mortality have ascribed values of $k_C$= 0.015 and $\mu$= 0.25, the diffusion coefficient D being allowed to assume various constant values.

### 2.2 Models 3 and 1 and solving method

Three models of the dynamics of the system are built, numbered in growing order of detail from 1 to 3. With Models 1 and 3, Eq.s 3-4 are solved by the classical FVM. Model 2 will be solved by Eq.s 6-7 arising from one modified form of the same method.

Model 3 incorporates the short range heterogeneities of the system in all detail. The spatial domain is discretized using control volumes that exactly fit the small rectangles of the system definition. In Model 1, the short range variable heterogeneities are replaced by local averages. In Model 2, the heterogeneities lose their specified locations and shapes, keeping whenever possible the real sizes, distances and boundary magnitudes.

Exact solution of Eq.s 1-2 do not exist for most choices of the parameter values, the initial and boundary conditions. Approximated methods are then used to simulate the evolution of the system. The finite volume method applied to the defining Eq.s 1-2 will transform these equations in a set of algebraic equations that can then be iteratively solved, in order to calculate the evolution of the system as described by particular models. The finite volume method that will be used, originated at the team of Brian Spalding (see for example Artemov et al. 2009 and references there), is described in Patankar 1980, also in Versteeg et al. 1995.

In the FVM, the spatial domain is divided in control volumes where the terms of the equations are integrated, the diffusion flux term being further expressed in the form of a surface integral. The Eq.s 1-2 can be approximately expressed, for 2D domains, in the discretized algebraic forms of Eq.s 3-4.

$$I_P \left(\frac{1}{\delta t} - k_c S_P + \mu\right) = \frac{I_P^0}{\delta t} + \frac{A_w D_w}{\delta V \delta x}(I_W^0 - I_P^0) + \frac{A_e D_e}{\delta V \delta x}(I_E^0 - I_P^0) + \frac{A_s D_s}{\delta V \delta y}(I_S^0 - I_P^0) + \frac{A_n D_n}{\delta V \delta y}(I_N^0 - I_P^0) \quad (3)$$

$$S_P \left(\frac{1}{\delta t} + k_c I_P\right) = \frac{S_P^0}{\delta t} \quad (4)$$

The meaning of symbols can be found in the cited references. Between each pair of control volumes, the flux of one diffused quantity with density $\phi$ to the control volume $\delta V_P$ is $\frac{AD}{d}(\phi_n - \phi_P)$ where n is one neighbour control volume of P, A the area between them, D the diffusion coefficient and d the ascribed distance between the control volumes.

By iteratively solving Eq.s 3-4, the S and I fields of Model 3 are calculated in the space and time domains.

The explicit form adopted in Eq. 3 for the diffusive flux avoids I values at neighbour control volumes to be calculated at different times. The superscript 0 indicates values calculated at previous time iteration.

In Model 1, the spatial domain is discretized as to make each 3x3 control volumes defined in Model 3 to coincide with one control volume of Model 1, so having 79x39 control volumes. The initial S field is defined as the average of S in the 3x3 referred control volumes of Model 3. The initial uniform susceptible density for Model 1 is S= 100x1/9 + 20x8/9. The infected density is defined as I= 1/9 at the control volume centred at (x,y)= (150,1.923) and I=0 elsewhere.

The FVM is applied using the same Eq.s 3-4 as in Model 3. The variables used in the detailed fields of Model 3 are, in Model 1, replaced by spatially mean-field variables. This is accomplished just by the use of a coarse grid in which discrete variables are defined and calculated by means of the classical finite volume method. It is expected Model 1 to erase important details of the heterogeneous fields existing in both the system and Model 3.

### 2.3 Model 2 and solving method

In real cases, many of the heterogeneous details of one system may be unknown or hard to describe. The proposed Model 2 will use the coarse grid of Model 1, although preserving some of the heterogeneous features of the system and Model 3. Each control volume $\delta V$ is subdivided in a number of subvolumes $\delta V_K$ that may have specified or unspecified shapes and locations inside $\delta V$. Subvolumes belonging to the same control volume will have ascribed sizes, interface areas and distances between each pair of subvolumes, according to prior knowledge or guess. The finite volume method is then applied in a way similar to the original method.

The number of subvolumes inside each control volume, their sizes, the values of their interface areas and the distances between them must be quantified according to the existing knowledge and guesses about the system. These features may be very interesting in cases in which detailed geometry of the short range heterogeneities is not known or would imply excessive effort to be quantified. If optimized, the values found for these parameters may bring some new information about geometric characteristics of the heterogeneous configuration of the system.

Some modifications to the classical finite volume method are introduced to cope with the presence of $\delta V_k$ inside each $\delta V$. The reaction rates inside $\delta V_k$, the transport fluxes between different $\delta V_k$ of the same $\delta V$, and the transport fluxes between different $\delta V$, all have to be calculated. This last feature is associated with the exclusion of the flux calculation between $\delta V_k$ belonging to different $\delta V$.

Between each pair of subvolumes, $\delta V_k$, $\delta V_{k'}$, the flux of one transported quantity associated to the density $\phi$ is $\frac{A_{kk'}D}{d_{kk'}}(\phi_{k'} - \phi_k)$, with $A_{kk'}$ and $d_{kk'}$ the assigned contact area and distance between $\delta V_k$ and $\delta V_{k'}$, similarly to the flux between different control volumes. Reaction rates, now applied to the subvolumes, follow the same principles imposed by the FVM to the discretized versions of Eq.s 1-2, now applied to $\delta V_k$.

The control volumes $\delta V$ of Model 2 – in the present work identical to the ones of Model 1 – are subdivided in subvolumes $\delta V_k$. The average values of $\phi$ in $\delta V$ are here calculated proportionally to $\delta V_k$ by Eq. 5.

$$\bar{\phi} = \sum_k f_k \phi_k = \sum_k \frac{\delta V_k}{\delta V} \phi_k \qquad (5)$$

The discretized values of $I_k$ and $S_k$ of Model 2 at each location and time are calculated by Eq.s 6-7.

$$I_k \left(\frac{1}{\delta t} - k_c S_k + \mu\right) = \frac{I_k^0}{\delta t} + \sum_{k' \neq k} \frac{A_{kk'}D}{\delta V_k d_{kk'}}(I_{k'}^0 - I_k^0) + \sum_n a_n (I_{cvn}^0 - I_{cvP}^0) \qquad (6)$$

$$S_k \left(\frac{1}{\delta t} + k_c I_k\right) = \frac{S_k^0}{\delta t} \qquad (7)$$

$I_{cvn}$ in Eq. 6 are the values of I in the control volumes neighbour of $\delta V_P$. The subscript cv stands for control volume values which, in the present study, are calculated by means of Eq. 5.

In this method, the diffusion fluxes among subvolumes can be calculated once assumed the values of distances between them and the interface areas. The errors caused by limited knowledge of the geometry of spatial heterogeneities may be minimized by careful choice or optimization if possible of the values of $A_{kk'}/d_{kk'}$ and $\delta V_k$.

Other modelling choices being possible, the flux of I in Eq. 6 to one subvolume $\delta V_k$ is here split in two parts. One part calculates the flux from other $\delta V_{k'}$ belonging to the same $\delta V$, second term of right hand of Eq. 6. The other part do not directly seek the contributions of the $\delta V_{k'}$ inside other $\delta V$. Instead, it calculates the flux from neighbour control volumes to the $\delta V$ inside which $\delta V_k$ is located. The form of Eq.s 6 has implicit the supposition that the flux between control volumes $\delta V$ will be distributed among the subvolumes inside $\delta V$ proportionally to their spatial extent, Eq. 5, irrespective of the location of each $\delta V_k$ inside $\delta V$. The coefficients $a_n$ of Eq.s 6 and 3 are identical.

This submodel for the diffusion flux, corresponding to limited knowledge about system heterogeneity, obeys the necessary rule of conservation of extensive quantities. The number $\delta n_I$ of transported infected individuals between two different $\delta V$ is associated with the values of the intensive variable $I_{cv}$, the average infected densities, at both $\delta V$.

The present Model 2 will have 79x39 control volumes. The number of the supposedly unlocalized subvolumes forming each $\delta V$ is chosen by means of admitted evidence or by reasonable guess. In the present work, each control volume $\delta V$ of Model 2 is subdivided in two subvolumes, so k=1,2 in Eq. 6. With the particular description of the system in the present work, the two values of k may be supposed to represent the dual nature of the space domain at city and countryside, with disparate initial susceptible population densities.

All choices of the values of $A_{12}$ and $d_{12}$ compatible with $\delta V$ dimensions are possible. Suggested by the known specifications of the system, the distance between $\delta V_1$ and $\delta V_2$ is chosen to be $d_{12}=1.4$ and border length $A_{12}=5.1$, non-optimized values, nevertheless guided by similarities with Model 3, while $\delta V_1=\delta V/9$ and $\delta V_2=8\delta V/9$ carry the "true" values of the system.

Initial susceptible densities at all unlocalized $\delta V_1$ and $\delta V_2$ are respectively $S_1=100$ and $S_2=20$, while the infected densities are I=1 at the subvolume $\delta V_1$ of the control volume centred at (x,y)=(150,1.923) and I=0 for all the remaining subvolumes of all the control volumes of the spatial domain.

**2.4 Exact integral relations**

The option was made of impermeable borders. The flux of infected is null across external boundaries, as it was already for susceptible owing to the non-existent susceptible transport term.

The system of Eq.s 1-2 give rise, by space and time integrations, to Eq. 8 involving total numbers of infected and susceptible. Gauss theorem and non-permeability condition are used to determine this equation. Similarly obtained, Eq. 9 is valid at the time when the number of infected is maximum.

$$\int_{t_1}^{t_2} \mu n_I dt = -n_S(t_2) + n_S(t_1) - n_I(t_2) + n_I(t_1) \qquad (8)$$

$$\frac{\partial n_I}{\partial t} = 0 \implies \frac{\partial n_S}{\partial t} = -\mu n_I \qquad (9)$$

Equation 8 establishes one exact equality between functions of the total numbers $n_I$ and $n_S$. The numerical results for all Models of the present system, for all possible sets of constant coefficients, are expected to fit these criteria.

Eq.s 8-9 can be used to assess the precision of some integral results of the three Models, but not the overall quality of all results. Although of exact application to the defined system and the involved quantities, these relations do not avoid differences in the local spatial and temporal behaviour of other quantities, section 3.

Calculations are performed with time increments between $4\times10^{-3}$ at smaller D and $5\times10^{-5}$ at larger D values. With the time steps used in the present calculations, the explicit formulation of Eq. 3 has results that prove to be very close to the ones using the slower 4$^{th}$ order Runge-Kutta method, not presented in the next section.

## 3. Results

All three Models have initial I=0 except at the control volume approximately centred at (150,1.923). Zero transport flux is imposed at all the boundaries, what implies that lines of equal I tend to normal to the domain boundaries. Simulation results of the more detailed Model 3 are chosen as reference to check the reliability of the other two Models.

The evolutions of I field according to the three Models, Fig. 1, reveal the wave behaviour of the reaction-diffusion system. Waves of I and S (not shown) originate at the middle of the domain's lower side, where an initial non-zero I value is inserted, and propagate in all directions towards the remaining boundaries at constant speed. Figure 1a shows lines of equal infected density at equally separated times, using Model 3 with D=1, at t= 60, 120 and 180 time units.

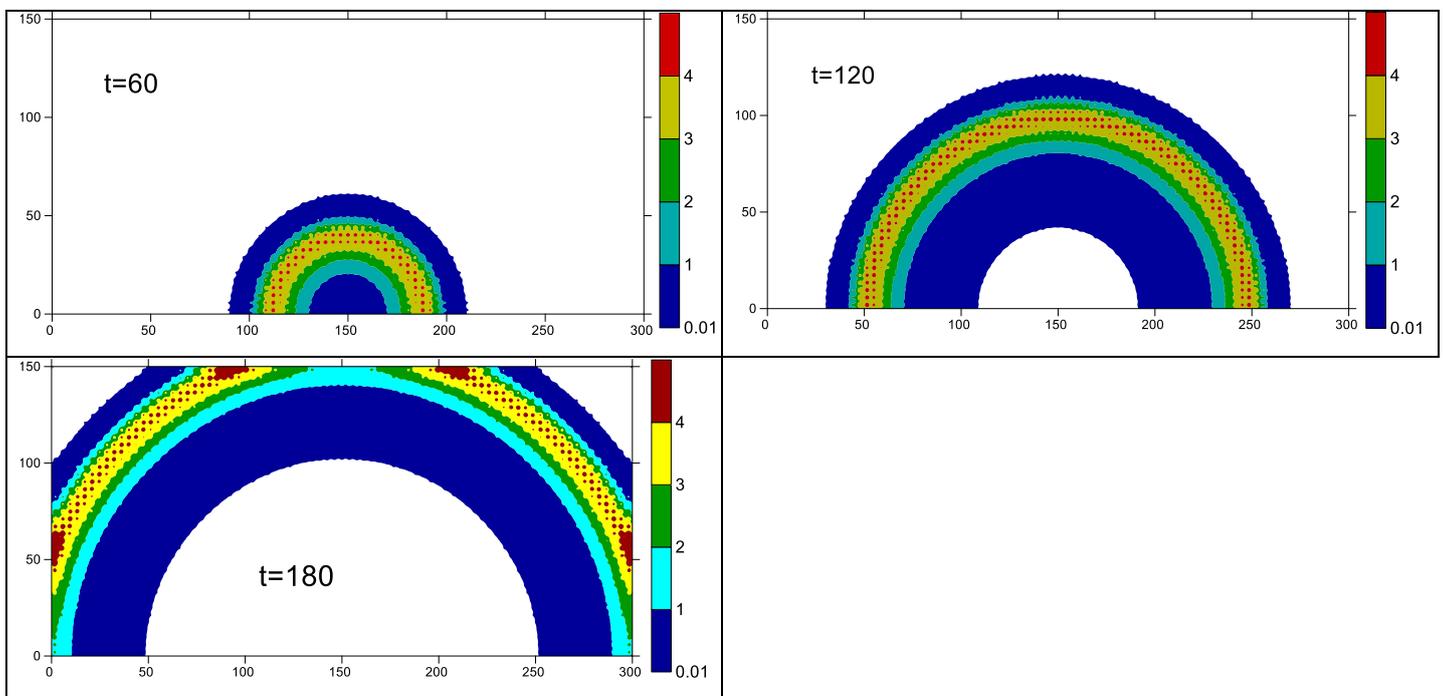

**Figure 1a.** Infected densities calculated by Model 3 with *D*=1 at different times.

In Fig. 1a wave fronts are circular far from the domain borders. S heterogeneities lead to I heterogeneities, apparent at locations with recent infections near the wave front. Far from the maxima of I wave, infected density is yet too low due to few infections, or already too low due to mortality.

At impervious borders diffusion only occurs parallel to the border, so leading to higher contagion rates and infected density. This is more evident at locations with longer intersection zones between the wave front and the border line (Fig. 1a, t=180). It is then expectable the final susceptible density at borders to be influenced by these features.

At larger D values (not shown in Fig. 1a) the infected heterogeneities become less sharp as effect of diffusion transport.

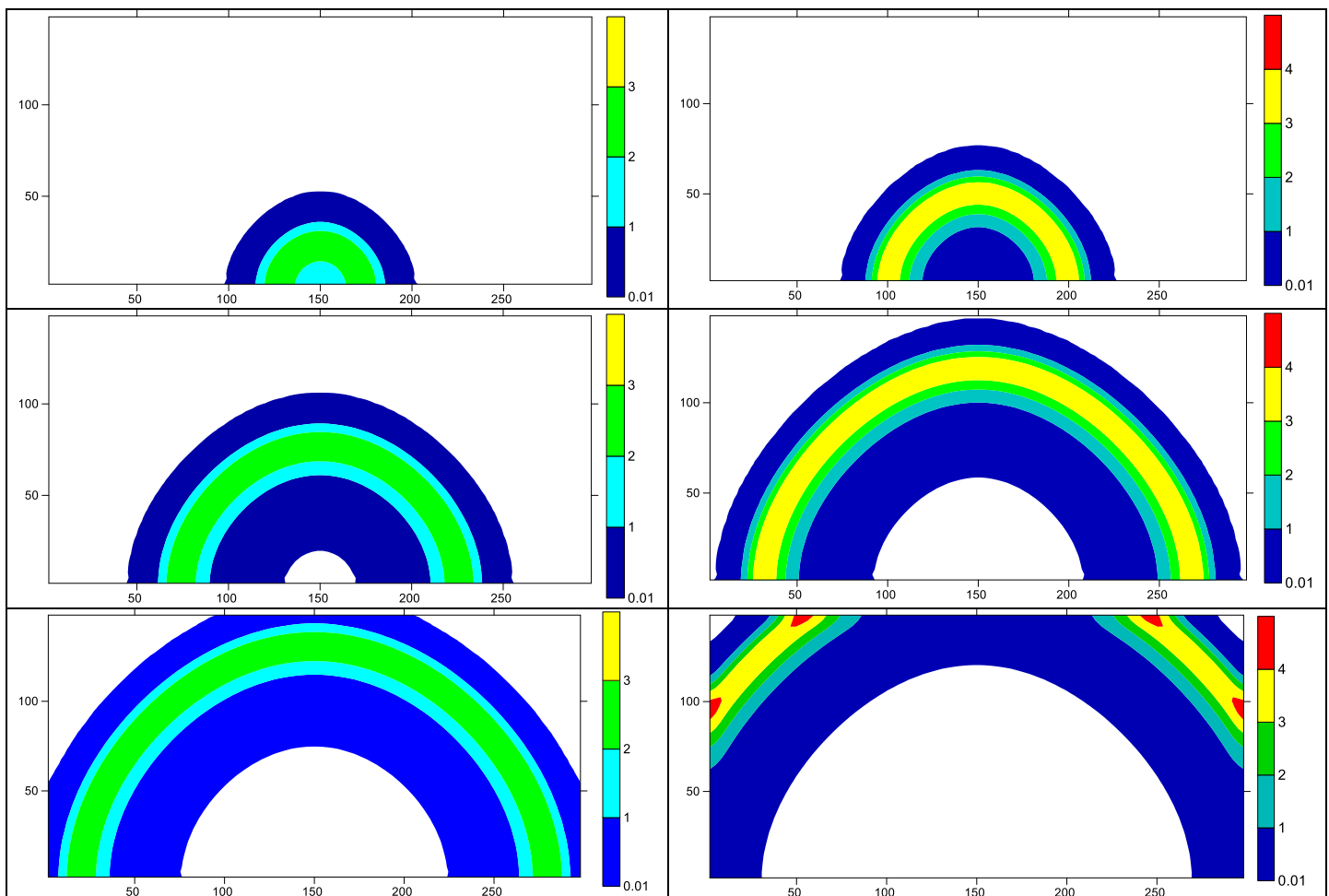

**Figures 1b-c.** Infected densities at times t=60, t=120, t=180, with Models 1 (left) and 2 (right) with D=1.

Using Model 2, Fig. 1c, t=180, high values of I are visible associated to the fact that the impervious walls inhibits the diffusive spreading of infected to outside the domain. The same is observed at Fig. 1a with the high detail specific of Model 3.

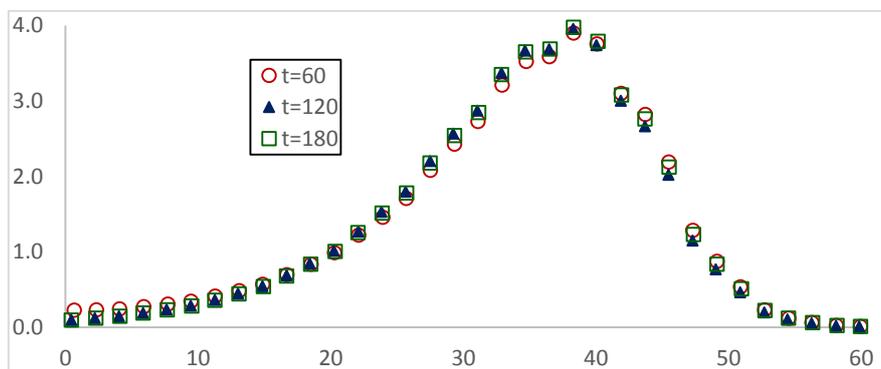

**Figure 2a.** Profiles of I along line **r**, see text, at t=60, 120 and 180, for Model 3, D=1, with translation of the profiles to make $I_{max}$ position to coincide with the peak at t=60.

Fig. 2a shows the I profile at the line **r** from (150,0) to (300,150), at different times calculated with Model 3. The value of $I_{max}$ remains constant at different times. At t=60, $I_{max}$ is located at r=38.3, while at t=120 and 180 $I_{max}$ is located at r=97.8 and 157.2 along line **r**. Making $I_{max}$ at later times to graphically coincide with the plot of t=60, Fig. 2a shows the wave form to be unchanged. The distance between $I_{max}$ at different t reveals near constant velocity of the peak.

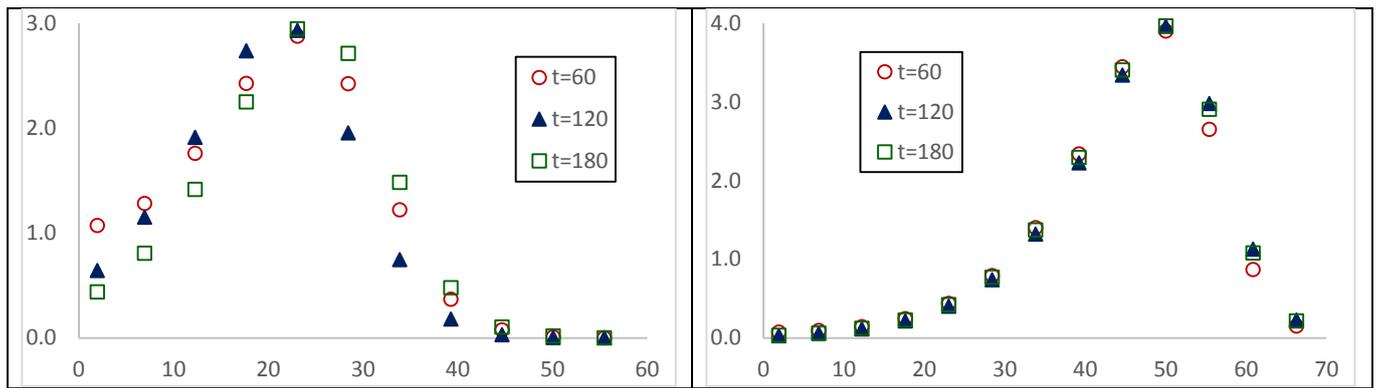

**Figure 2b-c.** Profiles of I along one line r at t=60, 120 and 180, for Models 1 (left) and 2 (right) with D=1, with translation of the profiles to make $I_{max}$ position to coincide with the peak at t=60.

Figures 2b-c show travelling waves with nearly unaltered profiles of I(r) for Models 2-3, waves propagating at near constant velocities for each Model, Table I.

Initial wave velocity, t<60 (not shown), with the present initial conditions, is slower than later. This is thought to result from initial spread to zones with yet quite low I and having larger curvature than later.

| t  M | 60 | 120 | 180 |
|---|---|---|---|
| 1 | 23.0 | 77.1 | 125.7 |
| 2 | 50.1 | 114.8 | 174.3 |
| 3 | 38.3 | 97.8 | 157.2 |

**Table I.** Positions of $I_{max}$ at different times for each Model.

The intrinsically heterogeneous Models 2 and 3 result in higher velocities than calculated by Model 1. This is thought to result from the non-liner reaction rate. In heterogeneous S fields of Models 2-3, the infection growth in high S locations is fast, resulting in greater I gradients, so leading to larger diffusion fluxes.

Using fixed µ and $k_C$, total numbers of infected and susceptible change in time as Fig.s 3 show at various D.

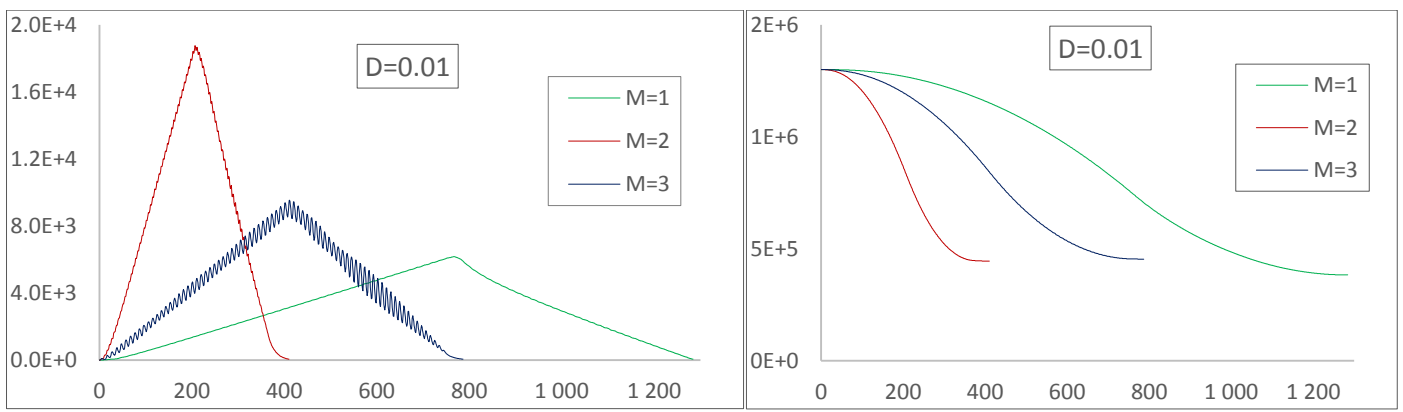

**Figure 3a.** Total infected (left) and susceptible with time.

At small D, the slow and thin I traveling wave has, in the case of Model 3, parts both in locations of small and large S, what produces oscillations, I(x,y,t) highly depending on time and place of observation.

Correspondingly, the results of Fig. 3a for $n_I(t)$ show very large oscillations. The present 2D study displays in Fig. 1a the heterogeneity of I field, clearly visible in the wave front.

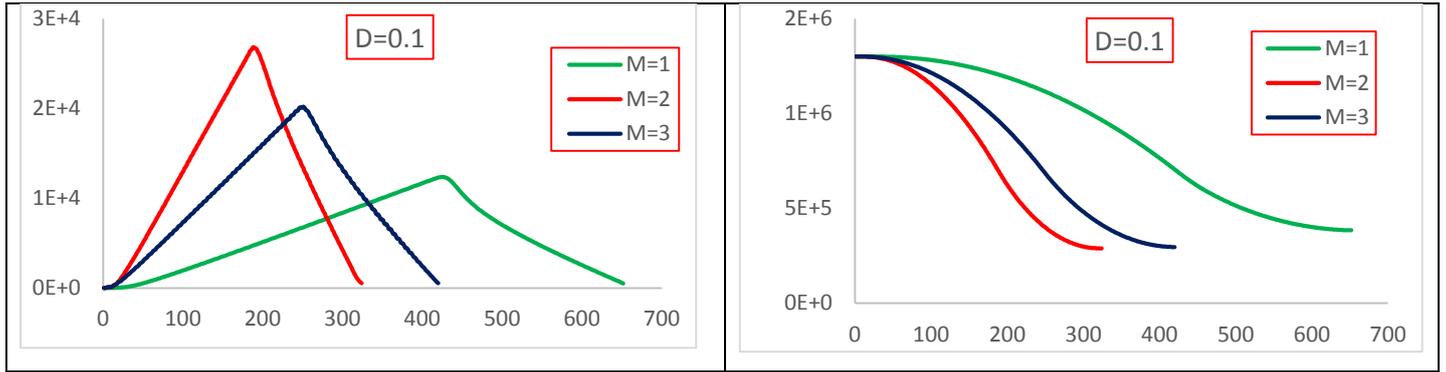

**Figure 3b.** Total infected (left) and susceptible with time.

From Fig.s 3, Model 2 compared to Model 3, exhibits at small D earlier and higher peaks of the total number of infected, the opposite occurring with Model 1.

The final numbers of susceptible is higher using Model 1 than using the other two Models at small and moderate D. The large death rates occurring in crowded areas of heterogeneous models, moreover causing large injection of infected in the neighbourhoods, has no exact homologous in areas with local uniform S characteristic of Model 1.

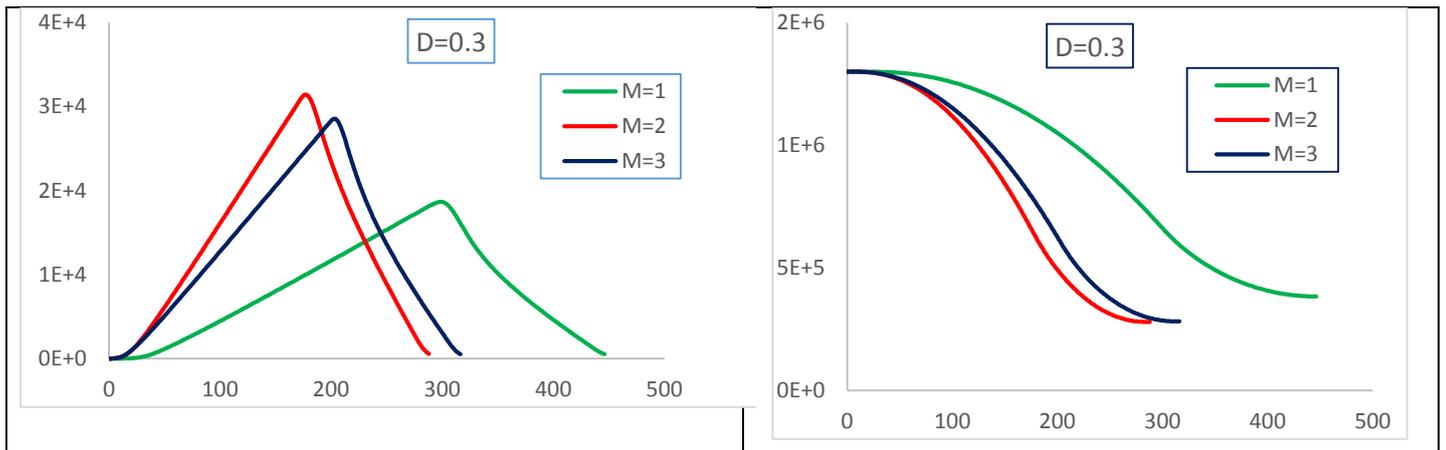

**Figure 3c.** Total infected and susceptible with time.

In Model 2, some geometrical details of Model 3 are lost, and diffusion between control volumes and between subvolumes depend on the chosen $d_{kk'}$ and $A_{kk'}$, also $\delta V_k$. With the used parameters, Model 2 leads, chiefly at small D, to faster calculated waves and a quicker infected rise. Larger values of maximum infected occur sooner than in Model 3 calculations, Fig.s 3a-c. At larger D, quasi-uniform values of I field have the effect of merging results.

At small and moderate D values, the maximum reached numbers of infected is low. As D increases, I propagation is faster, the maximum total number of infected grows and happens sooner, mortality is more intense as indicated by the disappearance of infected and the stabilization of the number of susceptible.

The total calculated mortality is smaller when using Model 1, although converging to the values obtained by Models 2 and 3 as D rises. The linear average of initial S used in Model 1 results in one lower average contagion rate, lower I flux, the propagation velocity being lower when Model 1 is used.

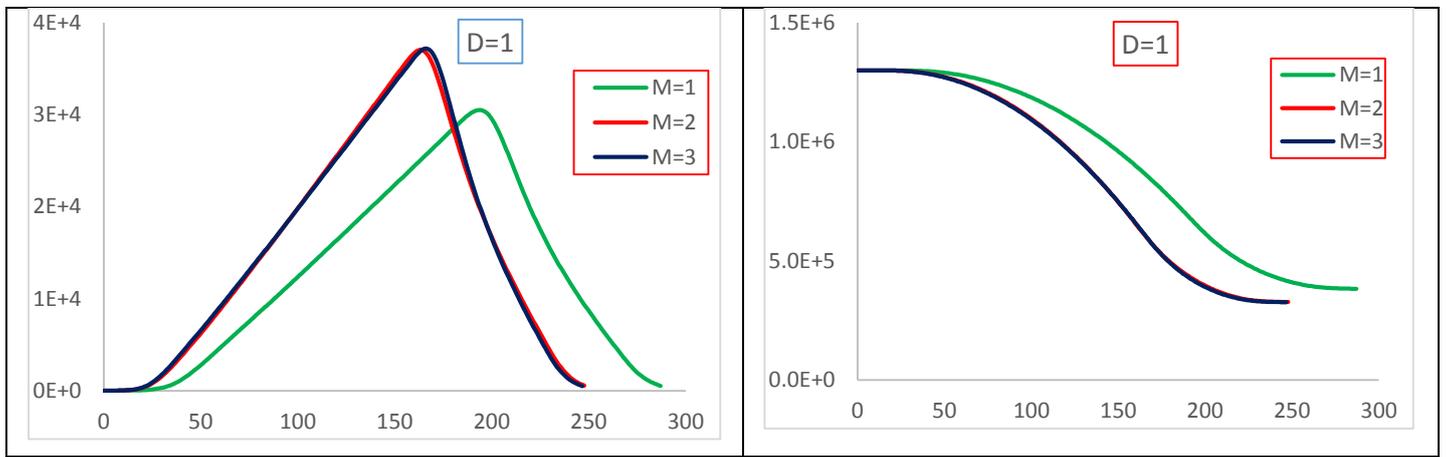

**Figure 3d.** Total infected and susceptible with time.

At large diffusion values, all the three Models show infected and susceptible numbers practically coincident, as result of the equalizing effect of the spread of I resulting from large diffusion fluxes.

Differently from what happens at small D values, at large D the substantial rise of $n_I$ is observed later. The intense spreading of infected into large areas not initially leads to important I densities growth. Larger rates of contagion than those at small D values arise later for large D. Some similarity may be found in the dynamics of many different systems, as in the case of well-stirred mixture of reactive fluids.

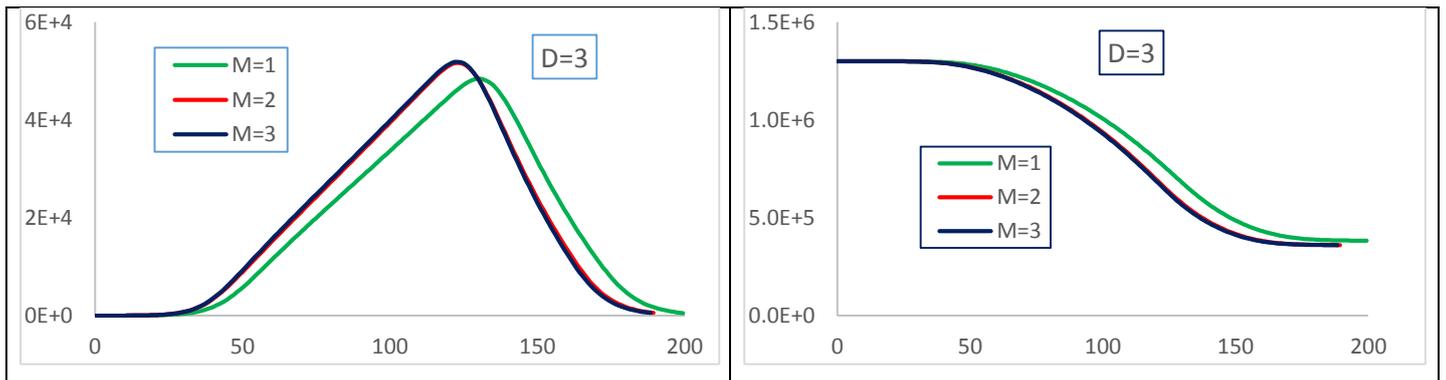

**Figure 3e.** Total infected and susceptible with time.

For D=1 the results of Models 2 and 3 are almost coincident. For values of D>10 the results of M=1 come to be close to the results of the other two Models, there offering increased reliability to the results of all the Models.

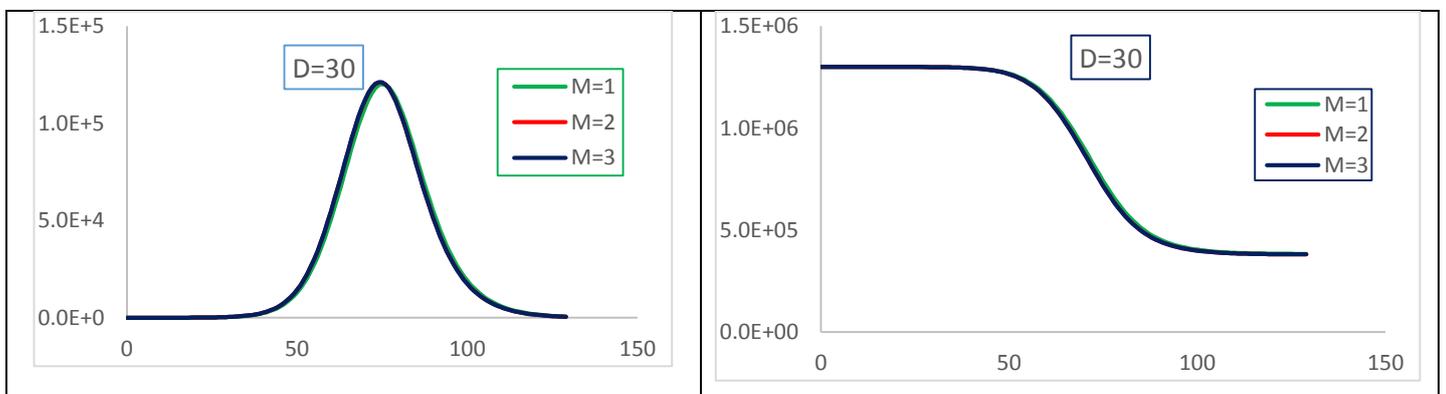

**Figure 3f.** Total infected and susceptible with time.

The final total values of the susceptible is nearly constant for Model 1, while for Models 2 and 3 increase with increasing D, slowly approaching the constant value around $n_S$=380000 (Fig.s 3 and 6).

Model 1, with erased local heterogeneities, has no extra contagion pumping caused by high density gradients existing in Models 2 and 3, no extra diffusion flux of infected existing from locations with large reaction rates. Therefore lower peak values of infected arise at later times than with the other two Models.

The calculated maximum total numbers of infected and the times at which they occur, available at Fig. 3, are synthetized in Fig.s 4-5 as a function of D.

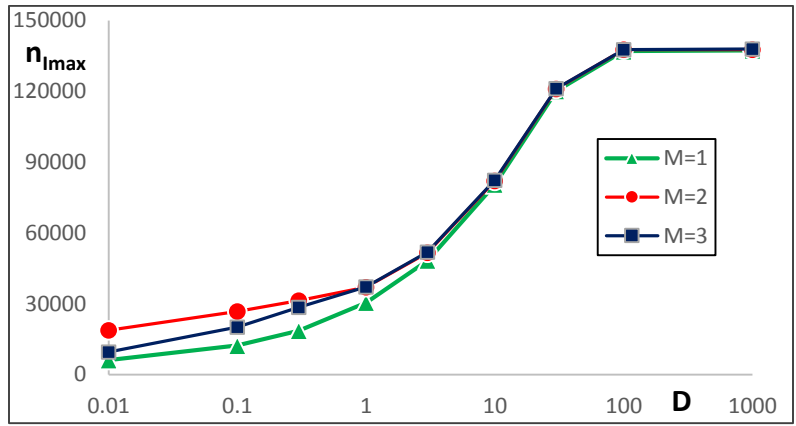

**Figure 4.** Maximum number of infected population as a function of diffusion coefficient.

Among all Models, the instant of the maximum number of infected clearly differ for D≤1 and almost coincide for D>3. For small values of the diffusion coefficient, the peak values of the total infected calculated by Model 2, with its chosen parameters, are larger than using Model 3 and occur sooner (Fig.s 4-5). The opposite happens with Model 1, the maximum number of infected being smaller and occurring later at small D (Fig. 5), with larger deviations than Model 2 in relation to Model 3. As D increases, Model 2 results merge with Model 3 at lower D values than Model 1 does.

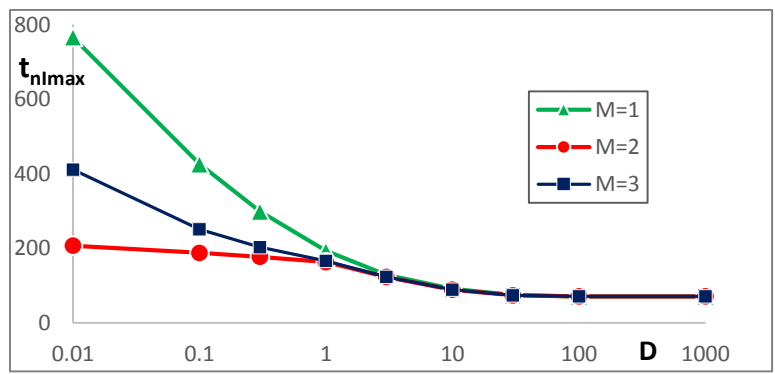

**Figure 5.** Times at which maxima infected population occur as a function of D.

The negative source term of Eq. 2 has the effect of a gradual $n_S$ decrease with time (Fig. 3). The total susceptible numbers then reduce asymptotically to final and minimum value exposed at Fig.s 3 and 6.

The final $n_S$ at extinction show a nearly fixed value for Model 1, what is not the case for Models 2 and 3. These two Models have very similar levels of final $n_S$ at different D, also shown in Fig. 3.

For Models 2-3, as D decreases to zero, $n_{Sfin}$ has one asymptote that exhibits high values of population survival. As D rises, Fig. 6 shows the decrease of $n_S$ to minimum values, maximum mortality, at $D \simeq 0.195$, then increasing as D rises, converging to the final $n_S$ values of Model 1 in one asymptote of lower $n_{Sfin}$ than at D→0. For real epidemic systems with alike behaviour the administrative decisions may be challenging.

Fig. 6 shows these final minima $n_S$ to depend on the values of the diffusion coefficient for the two Models in which heterogeneity is retained. On the other hand Model 1, that has no initial S heterogeneity, displays final $n_S$ practically independent of D.

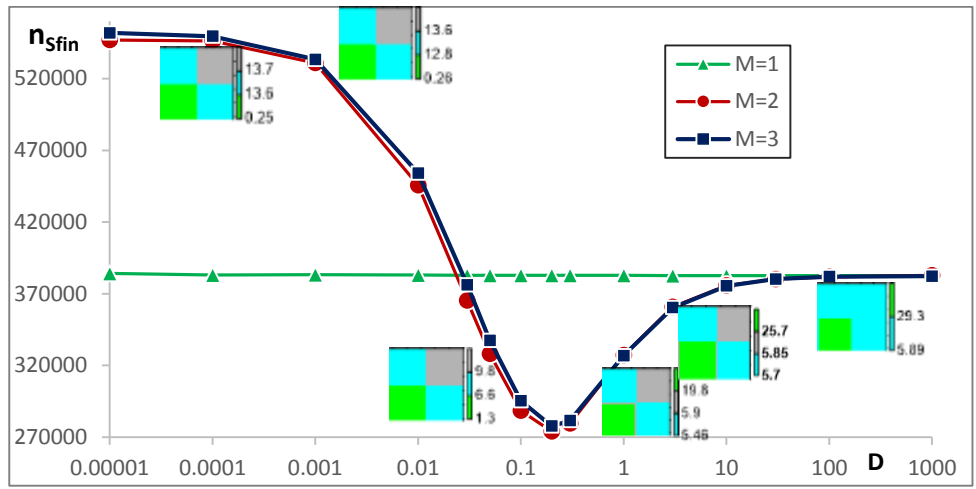

**Figure 6.** Number of susceptible at the end of the epidemic process for various diffusion coefficient values. The $S_{fin}$ densities displayed in the small figures refer to Model 3 at sites identified in Fig. 7, $D$=0.195.

At very small D≠0, at sites with high S, the high created I hardly travels to vicinities. The transported I to low $S_{ini}$ sites will have scarce time to infect S. Eq. 8 then points to not so many $n_{Sini}$ transforming in $n_{Sfin}$.

At high D, the large diffusion leads to nearly homogeneous I. The reaction rate in Eq. 2 so has equal values everywhere although varying in time. The fraction of S equally decreases everywhere for all Models.

Fig. 7 displays part of the final S field, $S_{fin}(x,y)$, at various D calculated by Model 3. Concerning S(x,y) of Model 3, three different types of control volumes are defined to clarify the characteristics of the numerical results. Control volumes **A** have $S_{ini}$=100. Among the control volumes with $S_{ini}$=20, those that share one boundary with one **A** site are labelled **B**, while those that share no boundary with **A** are labelled **C**.

The results suggest that, at small values of D with low flux of I, the high values reached by I at sites **A** are deleterious to S. Diffusion of I is inhibited and most of S is infected at **A** sites, during time enough to also eliminate non diffused infected at **A** sites. Sites **B** and **C**, with low $S_{ini}$, create less infected, that then disappear, while S has not time enough to be largely reduced.

At low D, the created I in **A** sites only in a limited way diffuse to neighbour sites, so depleting the S species in sites **A**. At intermediate D, it is expected the high production of I in **A** sites to diffuse to neighbour sites, so occupying larger space extents where S species is converted to infected. At the intermediate D range, the balance between reaction and diffusion is such that all kinds of sites finish with similar final S with the lower average values. The large space extent of sites **B** and **C** significantly contributes to final $n_S$ depletion. At large D, generated I is quickly diffused and tends to uniform. The depletion of susceptible is then proportional to S everywhere, Eq. 2. This is the reason for the convergence to the results of Model 1, in which also applies the proportionality to S depletion rate, Eq. 2.

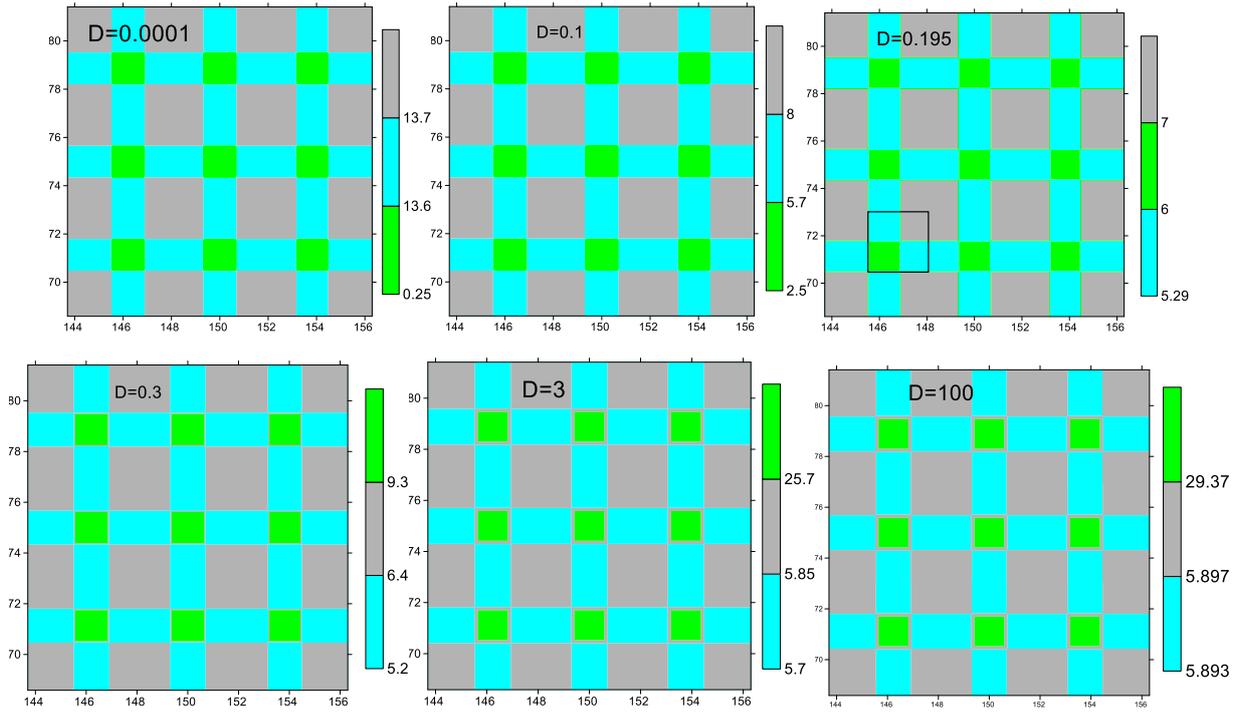

**Figure 7.** Final S field for various *D* values at typical sites of Model 3. At *D*=0.195 the window in Fig. 6 is displayed.

Table II displays $S_{fin}$ in the two subvolumes of Model 2, at the same D set as in Fig. 7. The results of Table II are quite similar to the averaged ones of Fig. 7.

| *D* | 0.0001 | 0.1 | 0.195 | 0.3 | 3 | 100 |
|---|---|---|---|---|---|---|
| $S_{fin1}$ | 0.252 | 6.86 | 6.04 | 9.3 | 25.8 | 29.26E |
| $S_{fin2}$ | 13.69 | 2.8 | 6.1 | 5.8 | 5.79 | 5.873 |

**Table II.** Final S values at the subvolumes of Model 2 at various *D* values.

In Fig. 7, green colour refer to sites **A**, with high initial S, irrespective of the final S values. For other sites, blue refers to relative low values, grey to high relative values. The results show that $n_{Sfin}$(**B**)< $n_{Sfin}$(**C**) at all D values simulated.

Moreover, only near D=0.2, where $n_{Sfin}$ is minimum, blue and grey colours have not consecutive values and change relative positions, with $n_{Sfin}$(**B**)< $n_{Sfin}$(**A**)< $n_{Sfin}$(**C**). This happens in the course of the transition of $n_{Sfin}$(**A**) from minimum to maximum as D rises.

At very small D=$10^{-4}$, $S_{ini}$=100 in δV of kind **A** will become $S_{fin}$=0.25. This remarkable variation may be interpreted in the context of very small I diffusion, so leading to the persistent reaction of S species with species I with very low flux away. Sites **B** and **C** retain relative high S values compatible with low I transport from sites **A**. Negative I source of Eq. 1 is enough to keep low I while reducing $S_{ini}$=20 to relatively high final values without catastrophically increase I values at these **B** and **C** sites.

As D rises at small and moderate values, $S_{fin}$ at **A**, **B** and **C** sites approach each other keeping $S_{fin}(A) < S_{fin}(B) < S_{fin}(C)$. Near $D \simeq 0.2$, $S_{fin}$ values are quite close, at higher D changing the relative order of $S_{fin}(A)$, $S_{fin}(B)$ and $S_{fin}(C)$, Fig. 7.

At higher D, type **A** sites become the final most populated control volumes calculated by Model 3. Figure 7 shows, at D=100 as at other large D, final S at **A**, **B** and **C** sites to approximately have the same fractional value of local initial S. At large D, the I field is nearly uniform, acting in the reactive term of Eq. 2 almost equally everywhere.

The fact that $S_{fin}$ of Model 1 is approximately equal at all D has one similar explanation. At almost all times and locations I is nearly uniform, leading solutions of Eq. 2 to equal values of $S_{fin} / S_{ini}$.

It may be expected that, at higher $\mu$, one fractional variation of S of similar amount would be attained, if $k_C$ and D are augmented proportionally to $\mu$. In fact, simulations with results not reported, using coefficients ten times larger, led to very close results, although occurring quite earlier than in the previous case, as expected from Eq.s 1-2.

## 4. Conclusions

An epidemic reaction-diffusion system has been defined, having short range heterogeneous initial S density. The system dynamics rules are established by means of balance equations of the species' densities.

The defined system is the base of three discrete models, more suitable to numerical simulations. Model 3 is a detailed representation of the heterogeneous system, now described by means of algebraic equations. Model 1 has been built, in which the short range heterogeneities have been converted to spatial averages of the density of the species.

Model 1 often oversimplifies the characteristics of the system, so Model 2 was introduced not including all the geometrical details of Model 3, nevertheless retaining important characteristics of the heterogeneous density fields.

Models 1 and 3 are solved by means of the standard FVM. Model 2 is solved by means of a modified version of the FVM in which the short range heterogeneities of variables are introduced in the algebraic equations. Non-localized subdomains $\delta V_K$ with undefined shapes are inserted inside the control volumes $\delta V$ of the standard finite volume method. The calculation of source terms is made inside the $\delta V_K$ in the way of the FVM. Transport fluxes are calculated among $\delta V_K \in \delta V$, to which the flux between different $\delta V$ is added and spread among $\delta V_K$ preserving conservation.

This method applied to Model 2 allows reasonable accuracy calculating non-linear heterogeneous reaction-diffusion systems. In case of unknown details, parameter optimization may possibly reveal some heterogeneity characteristics.

The simulations performed show the propagation of the epidemic starting at the contagion origin in the form of circular traveling waves. The wave speeds calculated by Models 2 and 3 have very similar constant values. Model 1 results display the circular wave to be some 15% slower than in the other two Models. At small D, the intrinsically heterogeneous Models 2 and 3 result in higher velocities than calculated by Model 1. This is thought to be caused by the non-linear reaction rate, resulting in greater I gradients, so leading to larger diffusion fluxes.

Model 3 calculations show sharp heterogeneities of I field, Fig. 1a, particularly in the areas adjacent to the epidemic wave front, as result of the reaction of I with the heterogeneous S field. At large values of D, the infected heterogeneities very soon disappear.

At impervious borders, diffusion only occurs parallel to the boundaries. The gradient of I normal to border tend to zero, so leading to higher infected density and contagion rates. This is more evident at locations with longer intersection zones of the wave front with the domain limits. At t=180 using Model 2, high values of I are visible associated to the fact that the impervious walls inhibits the diffusive spreading of infected to outside the domain. The same is observed in Fig. 1a with the extra detail associated to Model 3.

The numerical epidemic waves generated by the three Models have nearly fixed profile of the I(r,t) travelling wave front, as shown in Fig. 2, verifying the relation I(r,t)$\simeq$ I(r-v$\tau$,t-$\tau$), v being the velocity and $\tau$ one certain time interval.

The variation in time of the total numbers of infected, Fig. 3, shows the rising $n_I$ followed by one decrease till extinction. At high D, the initial rising of $n_I$ occurs later as a consequence of the weak initial local rise of I due to large diffusion, causing weak initial contagion rate. The number of susceptible $n_S$ starts with a total fixed number decreasing to non-zero values when $n_I$ tends to zero.

The results of Model 3 show important oscillations of the number of infected at small values of the diffusion coefficient, Fig. 3. At small D, the S field remains reasonably intact before the arrival of the narrow infected wave, being then reduced. The propagation of I wave in the heterogeneous S field originates considerable $n_I$ oscillations.

These oscillations reminds the modeller that, referring Model 2, the instantaneous diffusion from outer $\delta V$ into $\delta V_K$ will cause large diffusion too early. Modifications to the Model 2 are then needed that may improve results at small D.

Model 1, with erased local heterogeneities, has no extra contagion pumping caused by high density gradients existing in Models 2 and 3. Model 1 has then lower peak values of infected, that occur later, than the other two Models.

The calculated total number of susceptible at the end of the epidemic, Fig. 6, is approximately constant for Model 1 at all D. Models 2 and 3 predict, at small D values, high numbers of $n_{Sfin}$. At D$\simeq$0.2 $n_{Sfin}$ for M=2,3 come to minimum values roughly with half the $n_{Sfin}$ observed at very small D. At D>0.2, $n_{Sfin}$ rises and approaches one horizontal asymptote that coincides with the constant final $n_S$ calculated by Model 1.

For Model 3, in cases of small D, Fig. 7 shows at **A** sites the large $S_{ini}$ to become very small $S_{fin}$. A large fraction of $S_{ini}$ converts into I, that stays in **A** long enough to extremely diminish S due to the almost impervious sites at low D. Sites **B** and **C** receive from **A** insufficient I to greatly deplete S. The relatively small $S_{ini}$ only moderately converts in I. Sites **B** and **C** occupy the great majority of the spatial domain, one condition to the large $n_{Sfin}$ at small D, Fig. 6.

Fig. 7 shows that, at large D, the fraction $S_{fin}/S_{ini}$ in Model 3 is approximately the same in all sites of the space domain. This occurs as a consequence of the near uniform spread of I, which converts $k_C I$ in one uniform multiplier in Eq. 2, although varying in time. Density S diminishes proportionally to S, I continuing almost uniform (not shown).

The fact that $S_{fin}$ of Model 1 is approximately equal at all D has one similar explanation. At all locations and time $S_{fin}/S_{ini}$ is a constant. At large D, the uniformized I acts the same way everywhere, reducing S by the same fraction as in the case of Models 2-3 at large D.

The results of Model 3 show $n_{Sfin}(B) < n_{Sfin}(C)$ at all simulated D, possibly caused by the more intense I diffusion from **A** to **B**. Sites **B** have smaller average distance to one **A** site and share with it one common border, while **C** sites have no common border with **A** sites and are in average more distant of **A** than **B** sites do. So it is expectable that $S_{fin}$ at sites **B** have values closer to **A** sites than **C** sites do, what does not happen at roughly D> 0.2. At small D, final S fields keep the relation $S_{fin}(A) < S_{fin}(B) < S_{fin}(C)$ till values D≃0.2, at which Sfin(**A**), Sfin(**B**) and Sfin(**C**) have similar values, changing relative order at larger D values, Fig. 7.

Excepting at small D, the results of Models 2 and 3 are very similar, despite the fact that Model 2 has loosely defined geometry. Model 2 and the associated method of solution allow fairly good results, what may be valuable in case of limited knowledge of field heterogeneities. The optimization of the geometrical parameters of the method has not been tested, being possible that it can accomplish more accurate results. By means of parameter optimization, Model 2 can in principle be used to characterize types of heterogeneity of incompletely defined systems on the basis of the number and values of geometric parameters defined inside control volumes.

The different results at small D between Models 2 and 3 in Fig.s 3-5 suggest different behaviours of diffusion fluxes between the two Models. At small D, the smaller time at which $n_{Imax}$ is reached by Model 2 (Fig. 3 and 5) is compensated, in the integral of Eq. 8, by the larger value of $n_{Imax}$ (Fig.s 3-4), so reaching almost equal $n_{Sfin}$ between Models 2 and 3, see right hand side of Eq. 8 and Fig.s 3-4.

However, to correct these results of Model 2, see for example Fig. 3.b, lower $\frac{A_{kk'} D}{\delta V_k d_{kk'}}$ could be applied by changing the parameters. Lower D could be applied, violating the assumed conditions. Changing one of the $\delta V_K$, for example the $\delta V_1$, would imply to change $\delta V_2$ with uncertain results. The fact that the exact $\delta V_K$ of the present system have already been adopted this then is not the advisable way to ameliorate the method, so another interdiction in the present case. Apparently it remains the chance of altering $A_{kk'}/d_{kk'}$ or at least one of them. In the present case that is in principle one large risk as it would modify the results at intermediate D values, so probably breaking the coincidence existing between Models 2-3.

The author is persuaded that, in order to obtain better results using Model 2 at small D without disturbing the good results obtained at intermediate D, the correct choice under the scope of Model 2 consists in making modifications to the way diffusion between neighbour $\delta V$ is distributed among $\delta V_K$. Attention is needed to keep conservation of extensive quantities (as for example the transported $n_I$).